\documentclass[showpacs,prb,preprintnumbers,amsmath,amssymb,twocolumn]{revtex4-1}
\usepackage[latin1]{inputenc}
\usepackage{graphicx}
\usepackage{dcolumn}
\usepackage{bm}
\usepackage[usenames]{xcolor}

\begin{document}

\title{Memory matrix theory of the dc resistivity of a disordered antiferromagnetic metal with an effective composite operator}
\author{Hermann Freire$^{1}$}
\email{hermann\_freire@ufg.br}
\affiliation{$^{1}$ Instituto de Física, Universidade Federal de Goiás, 74.001-970, Goiânia-GO, Brazil}

\date{\today}

\begin{abstract}
We perform the calculation of the dc resistivity as a function of temperature of the ``strange-metal'' state that emerges in the vicinity of a spin-density-wave phase transition in the presence of weak disorder. This scenario is relevant to the phenomenology of many important correlated materials, such as, e.g., the pnictides, the heavy-fermion compounds and the cuprates. To accomplish this task,
we implement the memory-matrix approach that allows the calculation of the transport coefficients of the model beyond the quasiparticle paradigm. Our computation is also
inspired by the $\epsilon=3-d$ expansion in a hot-spot model embedded in $d$-space dimensions recently put forth by Sur and Lee [Phys. Rev. B \textbf{91}, 125136 (2015)], in which they find a new low-energy non-Fermi liquid fixed point that is perturbatively accessible near three dimensions. As a consequence, we are able to establish here the temperature and doping dependence of the electrical resistivity at intermediate temperatures of a two-dimensional disordered antiferromagnetic metallic model with a composite operator that couples the order-parameter fluctuations to the entire Fermi surface. We argue that our present theory provides a good basis in order to unify the experimental transport data, e.g., in the cuprates and the pnictide superconductors, within a wide range of doping regimes.
\vspace{0.3cm} 
\end{abstract}


\maketitle

\section{Introduction}

The ubiquitous ``strange-metal'' phase that appears in the cuprate superconductors near optimal doping \cite{Ong,Yoshida,Tyler}, the iron-based superconductors \cite{Matsuda,Analytis}, and heavy fermions compounds \cite{Coleman}, to name a few, is probably the most outstanding challenge in the field of strongly correlated systems and quantum criticality.  This metallic phase displays a universal linear-in-$T$ resistivity at intermediate temperatures. Since this phase clearly possesses no well-defined quasiparticle excitations at low energies, it represents a universal example of a non-Fermi liquid state that emerges in those correlated materials. Notwithstanding this fact, the underlying microscopic mechanism of such a state remains largely unknown up to this date \cite{McGreevy,Hussey}. One problem associated with this conundrum is related to the issue that it is very difficult, from a theoretical viewpoint, to write down a fairly ``realistic" model that describes a highly resistive metallic state with no quasiparticles whose momentum relaxation mechanism yields $\rho\sim T^{\alpha}$, such that $\alpha<2$. By contrast, conventional metals are normally described by the paradigmatic Landau's Fermi liquid theory, in which the resistivity follows the well-known scaling relation $\rho\sim T^{2}$ due to both umklapp scattering and disorder.

In this respect, a prominent theoretical proposal to describe a ``strange-metal'' phase consists of assuming the existence of a quantum critical point\cite{Hertz,Millis} (QCP) at $T=0$ that is responsible for generating such a state at finite temperatures \cite{Varma,Sachdev}. Experimental results in many compounds are by now quite numerous and provide a strong support to this perspective. Interesting possibilities of quantum critical points include spin-density-wave\cite{Chubukov1,Chubukov2,Metlitski,Efetov,Freire1} (SDW), charge-density-wave\cite{Castellani,Wang2} (CDW), pair-density-wave \cite{PALee,Agterberg,Fradkin,Freire2,Wang} (PDW), onset of various nematic orders \cite{Kivelson}, loop-current orders \cite{Varma2,Freire3}, fractionalized phases \cite{Chowdhury}, preformed excitonic pairs \cite{Pepin}, among many others (see, e.g., \cite{Eberlein}). We note in passing that recently an evidence of a novel PDW phase in the cuprate superconductor Bi$_2$Sr$_2$CaCu$_2$O$_{8+x}$ has been given by Hamidian \emph{et al.} \cite{Hamidian} using scanned Josephson tunneling microscopy, which was in fact anticipated by the microscopic theories in the Refs. \cite{Freire2,Wang}. In the present work, we shall focus only on the question of a SDW quantum phase transition underlying the phase diagram of a given correlated material in order to assess specifically what is the corresponding effect on the dc electrical resistivity as a function of doping of the adjacent metallic phases of the system at intermediate temperatures. 

The study of the SDW quantum criticality has quite a long history in the field (see, e.g., \cite{Pines,Chubukov1,Chubukov2,Metlitski,Efetov,Freire4}), but we will not detail all those works here. In this description, the microscopic mechanism of the Cooper pair formation is associated with the exchange of short-range antiferromagnetic spin-density-wave (SDW) fluctuations \cite{Pines}, which can be enhanced in the vicinity of a SDW quantum critical point. From a numerical viewpoint, it has been recently established\cite{Schattner} via a sign-problem-free Quantum Monte Carlo approach that the paradigmatic spin-fermion model, originally proposed by Abanov and Chubukov \cite{Chubukov1}, indeed describes a high-$T_c$ dome-shaped superconducting phase with the correct $d$-wave symmetry, in agreement with the experimental situation. Despite this statement, we note that in the underdoped regime of the hole-doped cuprates it has been argued \cite{Kochetov,Sachdev2} that the spin-fermion model must be altered to account for the non-double occupancy constraint that should be enforced for this case. From a weak-to-moderate coupling perspective, many analytical works have pointed out that the Abanov-Chubukov spin-fermion model essentially flows to strong-coupling at low energies \cite{Metlitski}, and one has to be very careful in devising new approximate perturbative schemes to calculate its physical properties. In this sense, an ingenious proposal consists of the $\epsilon=3-d$ expansion within a hot-spot model embedded in $d$-space dimensions recently developed by Sur and Lee \cite{SSLee}. Interestingly, they obtain in their work a stable non-Fermi liquid fixed point at low energies that is perturbatively controlled near three spatial dimensions. 

\begin{figure}[t]
\begin{center}
\includegraphics[width=3.7in]{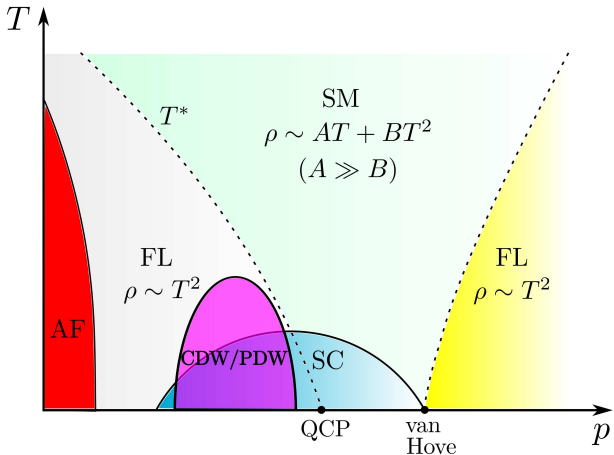}
\caption{(Color online) A possible phase diagram for the cuprate superconductors obtained in the present work based on the analysis of the dc resistivity of the two-dimensional spin-fermion model with both an effective composite operator and weak disorder. 
AF, SC, FL, and SM stand, respectively, for antiferromagnetism, $d$-wave superconductivity, Fermi-liquid-like behavior and strange metal phase. The QCP denotes the putative quantum critical point that is assumed to exist under the superconducting dome and
van Hove refers to the situation of van Hove band-filling in the model. For completeness, we also show the $d$-wave CDW/PDW composite order that is expected to manifest itself as a short-range order in the model.}\label{Phase_diagram}
\end{center}
\end{figure}

Transport calculations of the two-dimensional spin-fermion model are also of interest and have been performed by several researchers in the field \cite{Hartnoll,Patel,Patel2,Maslov,Efetov2,Maslov2,Pepin2}. Since it has become increasingly clear in the last years that there are no long-lived quasiparticles near the so-called ``hot spots'' (i.e., the points in reciprocal space where the antiferromagnetic boundary intersects the underlying Fermi surface) in the spin-fermion model \cite{SSLee}, a strong emphasis has been put on analytical approaches that allow the calculation of transport coefficients of this model without making any reference to the quasiparticle picture. One such approach is the Mori-Zwanzig memory matrix formalism \cite{Forster,Rosch,Hartnoll3,Freire5,Hartnoll2,Lucas}. The memory matrix is a generalization of the concept of scattering rate in Boltzmann theory and, for this reason, it can also be applied to strongly-correlated models describing non-Fermi liquid states in which this quantity is clearly not well-defined. In this respect, we point out that two recent papers \cite{Patel,Patel2} have applied this approach in order to investigate only the contribution to the resistivity due to the ``hot spots'' on the Fermi surface of the model. Consequently, they have shown that the resistivity contribution associated with those isolated points on the Fermi surface leads naturally to a constant term and a correction that is linear in temperature as a result of disorder. 

However, given the strong-coupling nature of the problem of SDW quantum criticality in two dimensions \cite{Metlitski}, it is conceivable that the order-parameter (bosonic) fluctuations of the spin-fermion model might ultimately couple not only to the aforementioned ``hot spots'', but crucially also to the remaining parts of the underlying Fermi surface of the model. Technically speaking, this proposal was first suggested by Hartnoll \emph{et al.} in Ref. \cite{Hartnoll}. In that work, they have shown that this can be achieved via the addition of a composite operator to the original model, which has the important effect of making the rest of the Fermi surface at least ``lukewarm'' (i.e., strongly renormalized), rather than simply ``cold'' (i.e., weakly renormalized) as is conventionally assumed in many works \cite{Hlubina}. 

Recently, Weiß \emph{et al.} \cite{Schmalian} were able to demonstrate microscopically the existence of such a composite operator explicitly from the spin-fermion model via an intermediate off-shell state. For this reason, we shall not repeat their derivation here. In other words, we assume, from the outset, the existence of such an effective term in the Lagrangian of the model. This higher-order interaction can be described in terms of a scattering process off a composite mode that includes, e.g., two spin fluctuations, such that the momentum transfer of fermions in the vicinity of the Fermi surface can be effectively small.  However, since this interaction involving the ``lukewarm'' fermions does not contain explicitly umklapp processes, it turns out that such a composite operator alone cannot render the dc resistivity of the model finite. Therefore, it is crucial to include other sources of momentum relaxation (e.g., disorder, phonons, etc) in the model on top of the composite interaction for the corresponding resistivity to become finite. Physically speaking, at high temperatures, the contribution from phonons is of course expected to play an important role in the temperature dependence of the resistivity of the present system. However, as the temperature is lowered, the impurity contribution will surely become the dominant one. For this reason, we shall focus here, as a first step, on impurity scattering alone as the main mechanism for momentum relaxation in the spin-fermion model with composite scattering, and leave the interesting analysis of the phonon contribution at higher temperatures for a future work. Thus, two significant questions, which one may legitimately ask at this point, are the following: What is the interplay of such a quantum critical composite interaction term with the effects of weak disorder on the transport properties (e.g., electrical resistivity) of the spin-fermion model at intermediate temperatures? And, secondly, can it possibly lead to an additional (perhaps, stronger) contribution to the resistivity of the model exhibiting the hallmarks of non-Fermi liquid scaling at finite temperatures with the correct doping dependence as observed experimentally? These are precisely the questions that we intend to address in the present work.

Our findings in the present work can be summarized pictorially (e.g., in the context of the cuprate superconductors) in Fig. 1. The two-dimensional model analyzed here describes a ``strange-metal'' phase at intermediate temperatures, insofar as the dc resistivity is described by $\rho_{xx}(T)\sim AT+BT^2$ (for $A\gg B$ at optimal doping). Moreover, on a more phenomenological level, in case there is pseudogap formation in the model at low doping (various proposals for this mechanism exist in the literature -- see, e.g., \cite{Chowdhury,Pepin,Harrison}), the hot spot regions would become clearly gapped out. In this scenario, the dc resistivity of the model revealed from our calculation would recover a traditional Fermi-liquid-like scaling given by $\rho_{xx}(T)\sim BT^2$. This result is surprisingly consistent with many recent transport experiments performed, e.g., in the cuprate superconductors  \cite{Greven,Mirzaeia,Abdel-Jawad}.

Our paper is organized as follows. First, we define the spin-fermion model with an effective composite mode in the presence of weak disorder. Secondly, we explain the formalism of the Mori-Zwanzig memory matrix approach in order to calculate transport properties of the model beyond the quasiparticle paradigm. Then, we move on to discuss our main results in this work and to compare them with previous results obtained in the literature. Lastly, we present our final conclusions.

\section{SDW Critical Theory with Composite Modes}

As previously explained for the SDW quantum criticality problem, our starting point will be the spin-fermion model\cite{Chubukov1,Chubukov2} with the inclusion of a composite operator\cite{Hartnoll,Abrahams} that effectively transfers small momenta and couples the order-parameter (bosonic) fluctuations to the whole Fermi surface of the model. In this way, the SDW critical field theory becomes described by the following Lagrangian

\begin{figure*}[t]
\begin{center}
\includegraphics[width=5.3in]{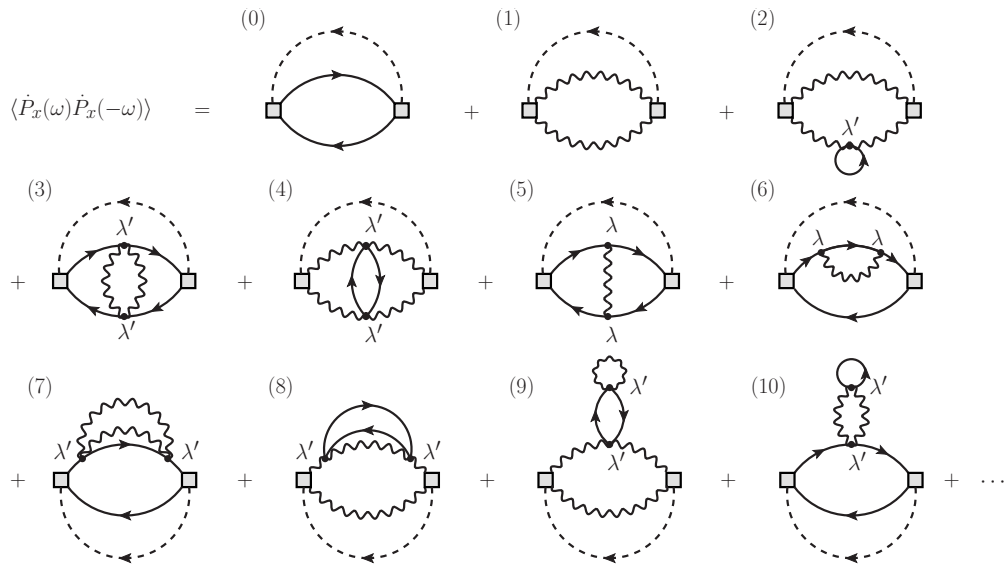}
\caption{Some Feynman diagrams that contribute to $G^{R}_{\dot{P_x}\dot{P_x}}(\omega,T)$ up to second order in the couplings. The solid lines refer to the fermionic propagators, while the wavy lines stand for the bosonic propagators. The coupling $\lambda$ denotes the inter-hot-spot scattering, whereas the coupling $\lambda'$ is the composite interaction that couples to the entire Fermi surface of the model. The impurity lines (dotted lines) only carry internal momentum and external bosonic energy $\omega$.}\label{Feynman_diagrams}
\end{center}
\end{figure*}

\vspace{-0.3cm}

\begin{eqnarray}\label{1}
\mathcal{L}&=&\sum_{\sigma}\bar{\psi}_{\sigma}(\partial_{\tau}+\bar{\varepsilon}_{\mathbf{k}})\psi_{\sigma}+\frac{1}{2}\chi^{-1}_{q}(\vec{\phi}.\vec{\phi})+\frac{u}{4!}(\vec{\phi}.\vec{\phi})^2\nonumber\\
&+&\lambda\sum_{\sigma\sigma'}\bar{\psi}_{\sigma}(\vec{\phi}.\vec{\tau}_{\sigma\sigma'}){\psi}_{\sigma'}+\lambda'\sum_{\sigma} \bar{\psi}_{\sigma}\psi_{\sigma}(\vec{\phi}.\vec{\phi}),
\end{eqnarray}

\noindent where $\chi_{q}^{-1}=(q_{0}^{2}+c^2\mathbf{q}^2+m)$ with the bosonic momentum $\mathbf{q}$ centered around the commensurate antiferromagnetic ordering wavevector $\mathbf{Q}=(\pi,\pi)$, $\bar{\psi}_{\sigma}$ and ${\psi}_{\sigma}$ are the fermionic Grassmann fields with spin projection $\sigma$, $\bar{\varepsilon}_{\mathbf{k}}={\varepsilon}_{\mathbf{k}}-\mu$ is the fermionic energy dispersion relative to the chemical potential $\mu$, $\vec{\phi}=(\phi_x,\phi_y, \phi_z)$ is the three-dimensional order-parameter collective field, $\vec{\tau}=(\tau_x,\tau_y, \tau_z)$ are the Pauli matrices, $c$ is the spin-wave velocity, $m$ is the bosonic mass that measures the distance of the theory to the antiferromagnetic QCP, $\lambda$ is the spin-fermion coupling constant and $\lambda'$ is the composite operator coupling constant. Regarding the last term in the Lagrangian (i.e., the composite interaction), it is worth mentioning that, by expressing it in momentum space, one can see that while the first spin fluctuation $\phi$ indeed carries the momentum $\mathbf{Q}+\mathbf{q}_1$ (where $|\mathbf{q}_1|\ll |\mathbf{Q}|$), the second order-parameter field $\phi$ must necessarily carry the momentum $\mathbf{-Q}+\mathbf{q}_2$ (where $|\mathbf{q}_2|\ll |\mathbf{Q}|$), such that the transferred momentum $\mathbf{q}_1+\mathbf{q}_2$ to the low-energy fermions is effectively small in the present theory.

In the next step, we proceed to introduce weak disorder in the model. We will add two types of disorder, which couple to different degrees of freedom in the present system, i.e., the short-wavelength disorder and the long-wavelength disorder. The former will couple directly to the fermionic fields in the model, while the latter will correspond to random shifts of the precise location of the SDW quantum critical point. Therefore, we must add to Eq. (\ref{1}) the following terms

\vspace{-0.3cm}

\begin{eqnarray}\label{2}
\mathcal{L}_{imp}&=&\sum_{\sigma}V(\vec{r})\bar{\psi}_{\sigma}(\vec{r})\psi_{\sigma}(\vec{r})+\sum_{\sigma}m(\vec{r})[\vec{\phi}(\vec{r}).\vec{\phi}(\vec{r})],
\end{eqnarray}

\noindent which should satisfy the Gaussian white-noise disorder averaging relations: $\langle\langle V(\vec{r}) \rangle\rangle=\langle\langle m(\vec{r}) \rangle\rangle=0$,
$\langle\langle V(\vec{r})V(\vec{r'}) \rangle\rangle=V_0^2\delta^2(\vec{r}-\vec{r'})$, and $\langle\langle m(\vec{r})m(\vec{r'}) \rangle\rangle=m_0^2\delta^2(\vec{r}-\vec{r'})$. Henceforth, we will refer to the parameter $V_0$ as a random potential for the fermionic field and the parameter $m_0$ as the random mass term for the bosonic field.

At this point, it is important to emphasize that the renormalization group scaling that is put forward by Sur and Lee in Ref. \cite{SSLee} for a hot-spot model embedded in $d$-space dimensions within a $\epsilon=d-3$ expansion implies the following power counting analysis: $[\omega]=\Lambda^{z_b}$, $[k_i]=\Lambda$,
$[u]=\Lambda^{3z_b-3+\epsilon}$, $[\lambda]=\Lambda^{(3z_b-3+\epsilon)/2}$, $[\lambda']=\Lambda^{(2z_b-3+\epsilon)}$, $[c]=\Lambda^{z_b-1}$,  where $\Lambda$ is a high-energy (ultraviolet) cutoff of the field theory, $z_b$ is the bosonic dynamical critical exponent, and  $i=x,y$ are the components of the momenta. As a result of this scaling, one finds straightforwardly a new stable non-Fermi liquid fixed point at low energies given by: $\lambda^*\rightarrow 20\pi\epsilon/3$, $(\lambda')^{*}\rightarrow 0$, $u^{*}\rightarrow 0$ and $z_{b}^*\rightarrow 1+5\epsilon/6$. We will assume here, as a first approximation, that at intermediate temperatures the effects of weak disorder will not change qualitatively the above fixed-point structure of the model in the clean limit. Another important remark here is related to the fact that this low-energy non-Fermi liquid fixed point is perturbatively controlled in the limit $\epsilon\rightarrow 0$, but, strictly speaking, not in the limit $\epsilon\rightarrow 1$. Fortunately, this is not a big limitation, in view of the fact that we will restrict our present analysis to an intermediate temperature regime such that $T>E_s$ (where $E_s\sim c^2\gamma$, with the parameter $\gamma\propto \lambda^2$ being the Landau damping constant of the model), i.e., before the fixed point of Sur and Lee has been reached.  We point out that this latter choice is also made on physical grounds, since at lower temperatures the above non-Fermi liquid fixed point is expected to be preempted by a $d$-wave superconducting phase, whose existence was recently confirmed via sign-problem-free Quantum Monte Carlo simulations of the spin-fermion model\cite{Schattner}. Therefore, in the intermediate temperature regime that we will consider here, we may assume, to a good approximation, that the spin-fermion coupling $\lambda$ lies within a perturbative regime, the composite operator coupling $\lambda'$ is small but finite, and the bosonic dynamical critical exponent is described by $z_b \simeq 1$ (i.e., the perturbative effects of Landau damping in the theory may be neglected). Finally, we also note the very recent works by Lee and collaborators \cite{SSLee2}, in which these authors were able to extend the above renormalization group results in the context of a non-perturbative calculation, and confirmed that $z_b=1$ at the low-energy non-perturbative fixed point of the spin-fermion model in $d=2$.

\section{Memory matrix approach}

Since the entire Fermi surface of the spin-fermion model described by the Lagrangian in Eq. (\ref{1}) can be potentially affected as a consequence of the aforementioned stable non-Fermi-liquid fixed point, it is absolutely necessary to have a methodology at our disposal that calculates transport properties without making any assumption with regard to the existence of quasiparticle excitations in our theory. Inspired by recent holographic methods applied to many fundamental problems in condensed matter systems (see, e.g., Ref. \cite{Holography}), a crucial realization emerged in the community that this can be achieved by using, e.g., the so-called Mori-Zwanzig memory matrix approach (for excellent explanations of this method in different contexts, see, e.g., Refs. \cite{Forster,Rosch,Patel,Lucas,Hartnoll2}). In this formalism, only the operators that are either conserved or nearly conserved, which have a finite overlap with the appropriate current of interest, are expected to play an important role in the computation of the transport coefficients. This is because such nearly-conserved operators have naturally the longest relaxation times in the theory. Therefore, these conservation laws turn out to be central to the memory-matrix approach and follow straightforwardly from the underlying symmetries of the Lagrangian. 

We start here with the matrix of generalized conductivities $\sigma(\omega,T)$ as a function of both frequency $\omega$ and temperature $T$, which can be written in this method as

\vspace{-0.3cm}

\begin{equation}
\sigma(\omega,T)=\frac{\chi^{R}(T)}{-i\omega+M(\omega,T)[\chi^{R}(T)]^{-1}},
\end{equation}

\noindent where $\chi^{R}_{AB}(T)=(A|B)=\chi^{R}_{AB}(\omega=0)$ is the matrix of the static retarded susceptibilities of some conserved (or nearly conserved) operators $A$ and $B$ in the system. This latter matrix is defined as 
$\chi_{AB}(i\omega,T)=\int_{0}^{1/T}d\tau e^{i\omega\tau}\langle T_{\tau}A^{\dagger}(\tau)B(0)\rangle$, where $\chi^{R}_{AB}(\omega)=\chi_{AB}(i\omega\rightarrow\omega +i0^{+})$, $\langle ...\rangle$ is the
grand-canonical statistical average, $T_{\tau}$ is the time-ordering operator, and the volume $V$ of the system has been set to unity. As for the memory matrix conventionally denoted by $M(\omega,T)$, it can be calculated in this formalism as follows

\vspace{-0.3cm}

\begin{equation}\label{4}
M_{AB}(\omega,T)=\int_{0}^{1/T}d\tau\left\langle \dot{A}^{\dagger}(0)Q\frac{i}{\omega-QLQ}Q\dot{B}(i\tau)\right\rangle,
\end{equation}

\noindent where $L$ is the Liouville super-operator, which is defined as $LA=[H,A]=-i\dot{A}$, with $H$ being the Hamiltonian of the system and $Q$ is a projection operator that projects out of a space of operators
spanned by all the conserved and nearly-conserved operators denoted by $\{A,B,...\}$. To simplify the memory matrix calculation, we have taken into account the fact that the nearly-conserved operators of the spin-fermion model have
the same signature under time-reversal symmetry in Eq. (\ref{4}), i.e., $(\dot{A}|B)=0$. The memory matrix is fundamentally related to the mechanism of relaxation of all nearly-conserved operators in the system. In addition to this fact, due to the presence
of various projection operations in Eq. (\ref{4}), this matrix is expected to be a smooth function of all coupling constants in the theory. Consequently, within a weak-to-moderate coupling regime, this quantity can be computed via perturbative means.

The Lagrangian defined by Eq. (\ref{1}) is invariant under both global $U(1)$ symmetry and spatial translation. As a result, using Noether's theorem, one finds that both electrical current $\mathbf{J}$ and the momentum operator $\mathbf{P}$ of the model are conserved at the classical level. In the present model, they are given by

\vspace{-0.3cm}

\begin{eqnarray}\label{5}
\mathbf{P}&=&\int d^2 x\left[i\sum_{\sigma}\nabla \bar{\psi}_{\sigma}\psi_{\sigma} + (\partial_t \phi) \nabla\phi \right],\nonumber\\
\mathbf{J}&=&-\frac{i}{m}\sum_{\sigma}\int d^2 x \nabla \bar{\psi}_{\sigma}\psi_{\sigma}.
\end{eqnarray}

\noindent At the quantum level, one expects that the corresponding operators will have the longest relaxation times and, for this reason, we will argue that they should dominate the transport properties in the present model. With the subsequent addition of weak disorder [Eq. (\ref{2})] to the Lagrangian describing the field theory, it is important to emphasize that the momentum operator does not conserve any longer due to the resulting breaking of translation symmetry. Thus, we obtain

\vspace{-0.3cm}

\begin{eqnarray}\label{6}
i\dot{\mathbf{P}}&=&\int \frac{d^2\mathbf{q}}{(2\pi)^2}\int \frac{d^2\mathbf{k}}{(2\pi)^2}\mathbf{k}\bigg[V(\mathbf{k})\sum_{\sigma}\bar{\psi}_{\sigma}(\mathbf{k}+\mathbf{q})\psi_{\sigma}(\mathbf{k})\nonumber\\
&+&m(\mathbf{k})\phi(\mathbf{q})\phi(\mathbf{-q-k})\bigg].
\end{eqnarray}

\noindent In the following, we shall use the propagator of $z_b= 1$ bosons at intermediate temperatures in the spin-fermion model at a critical doping (we set $c=1$)

\vspace{-0.3cm}

\begin{eqnarray}\label{7}
\chi(q_0,\mathbf{q})&=&\frac{1}{q_{0}^{2}+\mathbf{q}^2+R(T)},
\end{eqnarray}

\noindent where $q_0$ denotes the Matsubara bosonic frequency, $\mathbf{q}$ is the wavevector and we have neglected the spin-wave mass term from this point on. Additionally, following previous works  \cite{Chubukov3,Patel},
we have included a low-energy cutoff in the theory described by $R(T)$. This term was exactly calculated in Ref. \cite{Chubukov3}, and it was shown to be given by $R(T)=4\ln^2[(\sqrt{5}+1)/2]T^2$ for the case of the
dynamical critical exponent $z_b=1$. As for the fermions, we shall restrict
our analysis to the immediate vicinity of the underlying Fermi surface and linearize the energy dispersion as follows: $\bar{\varepsilon}_{\mathbf{k}}=\vec{v}_{\mathbf{k}}.\mathbf{k}+u'\mathbf{k}^2$, where $\vec{v}_{\mathbf{k}}$ is the Fermi velocity and $u'$
is related to the curvature of the Fermi surface.

Therefore, the dc resistivity $\rho_{xx}(T)=\lim_{\omega\rightarrow 0}[1/\sigma_{xx}]$ of the present model becomes given by

\vspace{-0.4cm}

\begin{eqnarray}\label{8}
\rho^{-1}_{xx}(T)&=& \Gamma^{-1}(T)\chi_{P_x J_x}(T),
\end{eqnarray}

\noindent where we have introduced the matrix of relaxation rates $\Gamma^{-1}(T)=\chi_{J_x P_x}(T)M^{-1}_{P_xP_x}(T)$. To leading order, the susceptibility $\chi_{J_x P_x}(T)$ is taken to be the noninteracting susceptibility that naturally evaluates to

\vspace{-0.3cm}

\begin{eqnarray}\label{9}
\chi_{J_x P_x}(T)&=&\frac{1}{m}\int \frac{d^2\mathbf{k}}{(2\pi)^2} k_x^2 \frac{[n_F(\bar{\varepsilon}_{\mathbf{k}})-n_F(\bar{\varepsilon}_{\mathbf{k+q}})]}{(\bar{\varepsilon}_{\mathbf{k}}-\bar{\varepsilon}_{\mathbf{k+q}})}
\nonumber\\
&\approx&\frac{m}{\pi} \mu(T=0)+O(e^{-\beta \mu}),
\end{eqnarray}

\noindent where $n_F(\varepsilon)=1/(e^{\beta\varepsilon}+1)$ is the Fermi-Dirac distribution, $\mu$ is the chemical potential and $\beta=1/T$ is the inverse temperature.
From the above expression, one can clearly see that $\chi_{J_x P_x}(T)$ has no temperature dependence. Consequently,
the $T$-dependence of the resistivity will come solely from the memory matrix $M_{P_x P_x}(T)$. For this reason,
we now move on to this calculation.

We now consider the effects of weak disorder to the memory matrix. For the reasons already explained before, we may assume that, at intermediate temperatures, the spin-fermion coupling $\lambda$ lies within a perturbative regime, the composite operator coupling $\lambda'$ is small but finite, and both the random potential $V_0$ and the random mass term $m_0$ are also small. Since the equation of motion of the momentum [Eq. (\ref{6})] is of order linear in the parameters $V_0$ and $m_0$, the leading contribution of $M$ turns out to be quadratic in the same parameters. Since we would like to keep only the dominant contribution to the Liouville operator, we shall also replace this operator by its noninteracting value ($L\approx L_0$). In addition to this, for the same reason, instead of using the full Hamiltonian in all grand-canonical averages, we will replace it by the noninteracting Hamiltonian. As a result, the leading contribution of the memory matrix in the spin-fermion model becomes

\vspace{-0.4cm}

\begin{eqnarray}\label{10}
M_{P_x P_x}(\omega\rightarrow 0,T)&=& \lim_{\omega\rightarrow 0}\frac{\text{Im}\,G^{R}_{\dot{P_x}\dot{P_x}}(\omega,T)}{\omega},
\end{eqnarray}

\noindent where $G^{R}_{\dot{P_x}\dot{P_x}}(\omega,T)=\langle\dot{P_x}(\omega)\dot{P_x}(-\omega)\rangle$ is the corresponding retarded Green's function within the Matsubara formalism at finite temperature $T$. The corresponding Feynman diagrams are displayed in Fig. 2.

\section{Results}

Here, we compute in an explicit way the memory matrix of the present model. In what follows, we stress that we will not take into account the Cooperon channel in the calculation. This contribution is generally expected to lead to the phenomenon of weak localization at low temperatures, which we will ignore in this paper for simplicity. Therefore, the perturbative contributions to the memory matrix in the present model turn out to be the ones depicted in Fig. 2. Henceforth, these terms will be referred to in the exact same order as they appear in this latter figure. Thus, 

\vspace{-0.4cm}

\begin{eqnarray}\label{11}
M_{P_xP_x}(T)=\sum_{i=0}^{10}M^{(i)}(T)+\cdots .
\end{eqnarray}

\noindent The zeroth order Feynman diagram stands for the lowest order coupling of short-range disorder to the fermion fields. Its leading contribution will come from pairs of hot spots $(i,j)$, such that $i\neq j$, that are connected by a large momentum transfer denoted by a vector $\vec{Q}_{ij}$ in momentum space. The diagram then evaluates to

\vspace{-0.3cm}

\begin{eqnarray}\label{12}
M^{(0)}=V_0^2\,\text{Im}\left\{\sum_{i,j,i\neq j}\frac{Q^{ij\,2}_x}{\omega}\int_{\mathbf{k},\mathbf{q}}\frac{[n_F(\bar{\varepsilon}_{\mathbf{q}})-n_F(\bar{\varepsilon}_{\mathbf{k+q}})]}{\omega+\bar{\varepsilon}_{\mathbf{q}}-\bar{\varepsilon}_{\mathbf{k+q}}+i0^+}\right\},\nonumber\\
\end{eqnarray}

\noindent where $\int_{\mathbf{k}}=\int d^2 \mathbf{k}/(2\pi)^2$ and the limit $\omega\rightarrow 0$ should be taken. Using the identity $1/(x+i0^+)=\mathcal{P}(1/x)-i\pi\delta(x)$, the above inter-hot-spot scattering term essentially yields a temperature-independent contribution, which is given by

\vspace{-0.3cm}

\begin{eqnarray}\label{12a}
M^{(0)}=-\sum_{i,j,i\neq j}\frac{{Q^{ij\,2}_x} V_0^2  \Lambda^2}{4\pi^3 |{\vec{v}_{i}}\times{\vec{v}_{j}}|},
\end{eqnarray}

\noindent where the ultraviolet cutoff $\Lambda$ must be imposed on the integrations over all the energies $\bar{\varepsilon}_{i\alpha}$ relative to the chemical potential in the present theory. Since the local curvature parameter $u'$ in the fermionic energy dispersion turns out to be irrelevant in the low-energy stable fixed point\cite{SSLee} discussed in Sec. II, we neglected, for simplicity, this curvature term in the above computation. We point out that we will use the same approximation in all contributions to the memory matrix that follow in this work. Finally, the above anisotropic equation for $M^{(0)}$ will naturally contribute to the residual resistivity $\rho_0$ that should appear in the model at low temperatures (e.g., if
the superconducting phase that exists in the model is suppressed by the application of an external field).

The next diagram in Fig. 2 refers to the lowest order coupling of long-range disorder to the bosonic modes. It eventually becomes

\vspace{-0.3cm}

\begin{eqnarray}\label{13}
M^{(1)}(T)&=&m_0^2\,\text{Im}\Bigg\{\int_{\mathbf{k},\mathbf{q}}k_x^2\int_{E_1,E_2}\pi^2\text{sign}(E_1)\text{sign}(E_2)\nonumber\\
&\times&(E_1^2-R(T))\theta(E_1^2-R(T))\theta(E_2^2-R(T))\nonumber\\
&\times&\left(\frac{1}{2\omega}\right)\frac{[n_B(E_2)-n_B(E_1)]}{\omega+E_2-E_1+i0^+}\Bigg\}
\end{eqnarray}

\noindent where $n_B(\varepsilon)=1/(e^{\beta\varepsilon}-1)$ is the Bose-Einstein distribution and $\theta(x)$ is the Heaviside step function. In the above derivation (and also in the calculations that follow in the present work), we use the spectral function $A(E,\mathbf{k})$ of the bosonic propagator, i.e.

\vspace{-0.3cm}

\begin{equation}\label{18}
\chi(k_0,\mathbf{k})=\int_{-\infty}^{\infty}dE \frac{A(E,\mathbf{k})}{ik_0 -E}
\end{equation}

\noindent where $A(E,\mathbf{k})=\frac{1}{2\alpha_{\mathbf{k}}}{[\delta(E+\alpha_{\mathbf{k}})-\delta(E-\alpha_{\mathbf{k}})]}$ with $\alpha_{\mathbf{k}}=\sqrt{\mathbf{k}^2+ R(T)}$. Additionally, we use also the following identities 

\vspace{-0.3cm}

\begin{eqnarray}\label{18a}
\int\frac{d^2\mathbf{k}}{\alpha_{\mathbf{k}}}[\delta(E+\alpha_{\mathbf{k}})&-&\delta(E-\alpha_{\mathbf{k}})]=-\pi\text{sgn}(E)\nonumber\\ 
&\times& \theta(E^2-R(T)),
\end{eqnarray}

\noindent and

\vspace{-0.3cm}

\begin{eqnarray}\label{18b}
\int\frac{d^2\mathbf{k}}{\alpha_{\mathbf{k}}}&k_x^2&[\delta(E+\alpha_{\mathbf{k}})-\delta(E-\alpha_{\mathbf{k}})]=-\frac{\pi}{2}\text{sgn}(E)\nonumber\\
&\times&(E^2-R(T))\theta(E^2-R(T)).
\end{eqnarray}

\noindent By calculating Eq. (\ref{13}), $M^{(1)}(T)$ yields a Fermi-liquid-like $T^2$-contribution to the memory matrix calculation, i.e.

\vspace{-0.3cm}

\begin{eqnarray}\label{13a}
M^{(1)}(T)\approx \left(\frac{1.77 m_{0}^2}{16\pi^2}\right)T^2.
\end{eqnarray}

\noindent A quick examination of the above result also reveals that the prefactor is isotropic and independent of doping in the model. This should be contrasted with the other contributions to the memory matrix calculated in this work [such as, e.g., Eq. (\ref{12a})], whose prefactors are anisotropic and dependent on doping, in view of the appearance of cross products of Fermi velocities in those expressions that are clearly related to the band structure. 

As for the Feynman diagrams given by $M^{(2)}$ and $M^{(8)}$ in Fig. 2 that represent corrections due to the renormalization of the bosonic propagator at one-loop and two-loop orders respectively, it is straightforward to show using Eqs. (\ref{10}) and (\ref{18}) that these terms evaluate to zero. The same result also holds to $M^{(6)}$, $M^{(7)}$, $M^{(9)}$ and $M^{(10)}$, which represent various corrections to the fermionic propagators in the model. In the calculation of the latter terms, since the fermionic propagators have independent energies given by $\bar{\varepsilon}_{i\alpha}$, expressions of the type $\int d\bar{\varepsilon}/(i\omega - \bar{\varepsilon})^n=0$ for $n\geq 2$ naturally appear, and the corresponding Feynman diagrams vanish as well. 

The fourth diagram in Fig. 2 is one of the leading vertex corrections to the resistivity with composite interaction $\lambda'$. As we have explained before, it couples to the whole Fermi surface of the model. Computing this term analytically for the present model is quite cumbersome, but it can be of course evaluated numerically. We obtain that it yields a leading contribution given by

\begin{widetext}

\begin{eqnarray}\label{14}
&&M^{(3)}(T)=-\text{Im}\bigg\{\frac{V_{0}^{2}\lambda'^{2}}{\omega}\int_{\mathbf{k},\mathbf{q},\mathbf{k'},\mathbf{q'}}k_x^2
\,\frac{1}{\beta^3} \sum_{k_0,q_0,k'_0}\frac{1}{[(k'_{0}+q'_{0})^2+(\mathbf{k'}+\mathbf{q'})^2+R(T)]}\frac{1}{[k_{0}^{'2}+\mathbf{k'}^2+R(T)]}\frac{1}{(i\omega+iq_0-\bar{\varepsilon}_{\mathbf{k+q}})}
\nonumber\\
&&\times\frac{1}{(i\omega+iq_0+iq'_0-\bar{\varepsilon}_{\mathbf{k+q+q'}})}\frac{1}{(iq_0-\bar{\varepsilon}_{\mathbf{q}})}\frac{1}{(iq_0+iq'_0-\bar{\varepsilon}_{\mathbf{q+q'}})}\bigg\}\approx\sum_{i,j,i\neq j}\sum_{m,n,m\neq n}\left(\frac{7.93 V_{0}^2 \lambda'^{2}}{256\pi^3 | \vec{v}_{i}\times \vec{v}_{j}| | \vec{v}_{m}\times \vec{v}_{n}|}\right)T^4.
\nonumber\\
\end{eqnarray}

\noindent where $k_0$ and $q_0$ stand for fermionic Matsubara frequencies, whereas $k'_0$
refers to a bosonic frequency and the limit $\omega\rightarrow 0$ should be taken. The pairs of indices $(i,j)$ and $(m,n)$ run over all points of the Fermi surface connected by a small wavevector $\mathbf{q'}$. 
As can be seen from Eq. (\ref{14}), the composite interaction $\lambda'$ gives a non-Fermi liquid contribution to the resistivity that will be analyzed more carefully below.

Like the previous calculation, the solution of the fifth Feynman diagram in Fig. 2, which is an important vertex correction to the model with composite interaction $\lambda'$ as well, turns out to be also somewhat lengthy. For this term, we obtain that the leading contribution is

\begin{eqnarray}\label{15}
&&M^{(4)}=-\text{Im}\bigg\{\frac{m_{0}^{2}\lambda'^{2}}{\omega}\int_{\mathbf{k},\mathbf{q},\mathbf{k'},\mathbf{q'}}k_x'^{2}
\,\frac{1}{\beta^3} \sum_{k_0,q_0,k'_0}\frac{1}{(ik_0+iq_0-\bar{\varepsilon}_{\mathbf{k+q}})}\frac{1}{(ik_0-\bar{\varepsilon}_{\mathbf{k}})}\frac{1}{[(\omega+q'_{0})^2+(\mathbf{k'}+\mathbf{q'})^2+R(T)]}\frac{1}{[q_{0}^{'2}+\mathbf{q'}^2+R(T)]}
\nonumber\\
&&\times\frac{1}{[(q'_0-q_0)^2+(\mathbf{q'}-\mathbf{q})^2+R(T)]}\frac{1}{[(\omega+q'_0-q_0)^2+(\mathbf{k'}+\mathbf{q'}-\mathbf{q})^2+R(T)]}\bigg\}\approx\sum_{i,j,i\neq j}\left(\frac{0.96 m_{0}^2 \lambda'^{2}}{256\pi^2|\vec{v}_{i}\times\vec{v}_{j}|}\right)\Lambda^2,
\nonumber\\
\end{eqnarray}

\noindent where the limit $\omega\rightarrow 0$ should be taken and the indices $i$ and $j$ run over all points of the Fermi surface connected by a small wavevector $\mathbf{q}$. Therefore, Eq. (\ref{15}) yields a temperature-independent contribution to the resistivity. Because of this property, this term will also contribute to the residual resistivity $\rho_0$ of the model. 

Lastly, the sixth Feynman diagram is the so-called Altshuler-Aronov-type vertex correction with inter-hot-spot interaction $\lambda$ to the memory matrix. We mention here that this contribution has been explicitly calculated previously in Ref. \cite{Patel}. We agree here with their result for the present case, in which the bosonic dynamical critical exponent $z\simeq 1$. In this way, the corresponding diagram evaluates to

\begin{eqnarray}\label{16}
&&M^{(5)}(T)=-\lim_{\omega\rightarrow 0}\text{Im}\bigg\{\frac{V_{0}^{2}\lambda^{2}}{\omega}\int_{\mathbf{k},\mathbf{q},\mathbf{k'}} Q^{ij\,2}_x
\,\frac{1}{\beta^2} \sum_{k_0,q_0}\frac{1}{[k_{0}^{'2}+\mathbf{k}^{'2}+R(T)]}\frac{1}{(i\omega+iq_0-\bar{\varepsilon}_{\mathbf{k+q}})}\frac{1}{(i\omega+iq_0+ik'_0-\bar{\varepsilon}_{\mathbf{k+q+k'}})}
\nonumber\\
&&\times\frac{1}{(ik'_0+iq_0-\bar{\varepsilon}_{\mathbf{k'+q}})}\frac{1}{(iq_0-\bar{\varepsilon}_{\mathbf{q}})}\bigg\}\approx \sum_{i,j,i\neq j}\sum_{\alpha,\beta}\left(\frac{0.001 V_{0}^2 \lambda^{2} Q^{ij\,2}_x
}{ | \vec{v}_{i\alpha}\times \vec{v}_{i\beta}| | \vec{v}_{j\alpha}\times \vec{v}_{j\beta}|}\right)T,
\nonumber\\
\end{eqnarray}

\end{widetext}

\noindent where we have slightly changed our previous notation by including new indices $\alpha$, $\beta$ that now run over only the \emph{hot spots} of the Fermi surface, which are connected by the large momentum transfer denoted by the wavevector $\vec{Q}_{ij}$. Therefore, the above result gives a non-Fermi-liquid linear-in-$T$ contribution to the resistivity. 

Using Eq. (\ref{8}), and collecting all the above expressions, we obtain that the dc electrical resistivity of the present model finally becomes
\vspace{-0.3cm}

\begin{eqnarray}\label{17}
\rho_{xx}(T)=\rho_0+AT+BT^2+CT^4,
\end{eqnarray}

\noindent where $\rho_0\propto [M^{(0)}(T)+M^{(4)}(T)]/\chi_{J_x P_x}^2(T)$, $A\propto M^{(5)}(T)/(T\chi_{J_x P_x}^2(T))$, $B\propto M^{(1)}(T)/(T^2\chi_{J_x P_x}^2)$, and $C\propto M^{(3)}(T)/(T^4\chi_{J_x P_x}^2)$. As can be seen, the residual resistivity of the model has therefore two independent contributions associated with different scattering mechanisms in the present model. As a consequence, our transport theory implies that even though the magnitude of the residual resistivity $\rho_0$ is not entirely related to the coefficient of the linear resistivity $A$, there are some important correlations among each other in the present effective model.

It is also interesting to note here that the non-Fermi-liquid $T^4$-contribution to the resistivity of the model turns out to be subleading with respect to the short-wavelength disorder contribution calculated in Eq. (\ref{16}). For this reason, Eq. (\ref{17}) may be further approximated to

\vspace{-0.3cm}

\begin{eqnarray}\label{23}
\rho_{xx}(T)\approx \rho_0+AT+BT^2.
\end{eqnarray}

\noindent A crucial point that we wish to stress here is that the $BT^2$-contribution to the resistivity is not necessarily subleading to the $AT$-term. The reason for this is that the two contributions come from different sources of momentum relaxation: the $AT$-term is associated with a hot-spot scattering mechanism off short-wavelength disorder, whereas the $BT^2$-contribution in turn requires long-wavelength disorder coupled to the bosonic order-parameter fluctuations in the model. In other words, our results add support to the point of view that the spin-fermion model may indeed describe a ``strange-metal'' phase at intermediate temperatures, but with a resistivity described by $\rho_{xx}\sim AT+BT^2$ (the relative magnitude of the prefactors $A$ and $B$ are such that $A\gg B$ at optimal doping).  In addition to this fact, we also would like to point out that, while the scattering rate associated with the coefficient $A$ in the resistivity is highly anisotropic and doping-dependent [see Eq. (\ref{16})], the scattering rate related to the coefficient $B$ turns out to be isotropic and independent of doping [see Eq. (\ref{13a})]. This feature of our present transport theory is remarkably consistent with recent experimental measurements performed in the cuprate compounds, such as, e.g., HgBa$_2$CuO$_{4+\delta}$ (Ref. \cite{Greven}) and Tl$_2$Ba$_2$CuO$_{6+\delta}$ (Ref. \cite{Abdel-Jawad}). It is also worthwhile to comment on that our theoretical prediction clearly agrees with a proposed phenomenology put forward by Hussey and collaborators in order to describe various aspects of the transport data of the cuprates in the literature \cite{Hussey,Hussey2}. In addition, we mention that some aspects of our present work also agree qualitatively with other transport theories, which include the cold-spot-model proposed by Ioffe and Millis \cite{Ioffe} and the nearly antiferromagnetic Fermi liquid model put forward by Stojkovic and Pines \cite{Pines2}.

Another important consequence of our theory is that, for the physical situation in which there is pseudogap opening in the antinodal directions of the Fermi surface as revealed by angle-resolved photoemission experiments (we mention here that several proposals for this mechanism exist in the literature -- see, e.g., \cite{Chowdhury,Pepin,Harrison,Freire1,Sachdev2}), the corresponding hot spots become gapped out and disappear altogether. Therefore, the resulting dc resistivity of the present model should recover a Fermi-liquid-like scaling described by $\rho_{xx}\sim T^2$ within this range of temperatures. This analysis is surprisingly consistent with recent state-of-the-art experiments performed by Mirzaeia \emph{et al.} \cite{Mirzaeia} and Barisic \emph{et al.} \cite{Greven} for the single-layer HgBa$_2$CuO$_{4+\delta}$ compound inside the pseudogap state, in which they provided strong evidence that both optical conductivity and electrical conductivity of this phase conform to Fermi-liquid-like scalings. 

An additional outcome of the present result is that, for doping regimes that lie beyond the situation of van Hove band-filling in the single-band model, the hot spots cease to exist. The reason is that, in this case, there is no intersection of the underlying Fermi surface of the model with the antiferromagnetic zone boundary any longer. Therefore, in such a highly overdoped regime, the resistivity of the model according to our theory should again recover a Fermi-liquid-like result $\rho_{xx}\sim T^2$ as a function of temperature. This is of course consistent with transport experiments performed in all cuprate materials, where it is observed that the normal state in the overdoped regime is indeed described by the Fermi liquid theory. To make evident this point and to emphasize other similarities with the physics of the high-$T_c$ cuprates, we depict pictorially a phase diagram from the point of view of transport of the two-dimensional spin-fermion model obtained in the present work together with recent theoretical results\cite{Freire2,Wang} related to this model in Fig. 1.

Lastly, we point out that a similar evolution of doping and temperature dependence of the resistivity has also been observed in the metallic phase of many iron-based superconductors [see, e.g., the compounds BaFe$_2$(As$_{1-x}$P$_x$)$_2$ in the Refs. \cite{Matsuda,Analytis} and Ba(Fe$_{1-x}$Co$_x$)$_2$As$_2$ in the Refs. \cite{Canfield,Efremov}]. Even though these are in some sense complicated materials that exhibit many bands at the Fermi energy level, our present result could also suggest that the spin-fermion model with an effective composite interaction may capture universal aspects of correlated metallic systems in the presence of strong antiferromagnetic fluctuations.

\section{Conclusions}

We have presented the computation of the dc resistivity as a function of doping and temperature of the metallic states that emerge due to the existence of a putative SDW quantum critical point in the presence of weak disorder. This analysis is relevant to the phenomenology of many important correlated materials. For this calculation,
we have implemented the Mori-Zwanzig memory-matrix approach that importantly does not rely on the existence of well-defined quasiparticle excitations in this model at low energies. 

As an application of the present transport theory, we have compared our predictions to the experimental situation in the cuprate superconductors. In the ``strange-metal'' phase, the dc resistivity evaluates to $\rho(T )\sim AT + BT^2$ at intermediate temperatures (for $A\gg B$, close to optimal doping), where the scattering rate related to the prefactor $A$ is non-universal and strongly doping-dependent, while the scattering rate related to $B$ is universal. We have also correctly obtained the temperature dependence of the dc resistivity of the system as measured experimentally in recent works \cite{Mirzaeia,Greven}, both inside the pseudogap phase and in the overdoped metallic regime, whose transport coefficients indeed conform to Fermi-liquid-like scalings. 

It will be clearly very important to analyze also the magneto-transport of the two-dimensional spin-fermion model with an effective composite operator using the present theory. For this reason, we plan to perform in a subsequent work a similar analysis for the Hall angle and the magnetoresistance of the model within the memory matrix formalism. From an experimental point of view, the cotangent of the Hall angle $\theta_{H}$ in the strange metal phase of the cuprates at optimal doping is characterized by a scaling law given by $\cot(\theta_{H})\sim T^2$, which is seemingly Fermi-liquid-like, despite the fact that the resistivity is given by $\rho(T )\sim AT + BT^2$ (for $A\gg B$) within the same temperature regime. Therefore, only an approach that allows the calculation of the transport properties of a model, which does not rely on the quasiparticle picture is capable of reproducing such an important result. Indeed, given our encouraging results obtained in this work concerning the doping dependence of the dc resistivity of a disordered nearly antiferromagnetic two-dimensional metal, we believe that the present framework may provide a good basis in order to unify all the available experimental transport data, e.g., in the cuprate superconductors and also in the closely related iron-based superconductors, within a wide range of doping and temperature regimes.

\acknowledgments

I would like to thank V. S. de Carvalho for his help with the figures in the present manuscript. Financial support from CNPq under Grant Number 405584/2016-4 is also acknowledged.

\end{document}